# Optical forces through the effective refractive index

JANDERSON R. RODRIGUES[1,2,*] AND VILSON R. ALMEIDA[1,2]

[1] *Instituto Tecnológico de Aeronáutica, CEP 12228-900, Vila das Acácias, São José dos Campos, SP, Brasil*
[2] *Instituto de Estudos Avançados, CEP 12228-001, Putim, São José dos Campos, SP, Brasil*
*Corresponding author: jrr@ita.br*

**The energy-based methods as the Dispersion Relation (DR) and Response Theory of Optical Forces (RTOF) have been largely applied to obtain the optical forces in the nano-optomechanical devices, in contrast to the Maxwell Stress Tensor (MST). In this work, we apply first principles to show explicitly why these methods must agree with the MST formalism in linear lossless systems. We apply the RTOF multi-port, to show that the optical force expression on these devices can be extended to analyze multiple light sources, broadband sources, and multimode devices, with multiple degrees of freedom. We also show that the DR method, when expressed as a function of the derivative of the effective index performed at a fixed wave vector, may be misinterpreted and lead to overestimated results.**

Optical (transverse gradient) forces between two adjacent dielectric structures, due to the overlap of the evanescent field of the guided modes, were proposed by Povinelli *et al.* [1]. The optical forces can be rigorously calculated using the Maxwell Stress Tensor (MST) formalism [2] or, alternatively, in linear lossless dielectric materials, it may be obtained directly from the device's dispersion relation as a function of one of the structure's degrees of freedom (gap) [1,3,4]. Due to its computational simplicity and physical insights, the latter method has been more used than the former [3,4]. Furthermore, the Dispersion Relation (DR) method can be further simplified to express the optical force as a function of the mode effective (refractive) index derivative, with respect to the gap, leading to two expressions: one where the derivative must be performed at a fixed wave vector [5], and another where it must be performed at a fixed angular frequency [6-8]. The latter expression version can also be obtained in an alternative formal manner, by using the recently developed Response Theory of Optical Force (RTOF) method, proposed by Rakich *et al.* [9,10]. Besides that, all these expressions must agree with the MST formalism [1,9].

In this letter, we apply first principles to explicitly show why these energy-based methods must agree with the MST formalism. We analyze a typical nano-optomechanical device to show that the DR method, expressed in terms of the effective index with derivative performed at a fixed wave vector, may overestimate the optical force if the correct transformations are not used, thus disagreeing with the MST. We also use the RTOF theory to extend the correct expression to more general cases.

For a nano-optomechanical system composed of pure dielectric materials, the conservation of linear momentum states that [2,11]:

$$\frac{d\mathbf{p}_{EM}(\mathbf{r},t)}{dt} + \frac{d\mathbf{p}_{ME}(\mathbf{r},t)}{dt} - \nabla \cdot \overset{\leftrightarrow}{\mathbf{T}}(\mathbf{r},t) = 0, \quad (1)$$

where $\mathbf{p}_{EM}$ is the electromagnetic momentum density, $\mathbf{p}_{ME}$ is the mechanical momentum (or the force) density, and $\overset{\leftrightarrow}{\mathbf{T}}$ is the Maxwell Stress Tensor (MST), which represents the momentum flux, given by [2,11]:

$$\overset{\leftrightarrow}{\mathbf{T}}(\mathbf{r},t) = \frac{1}{2}\left(\varepsilon_0 \mathbf{E}(\mathbf{r},t)\mathbf{E}^*(\mathbf{r},t) + \mu_0 \mathbf{H}(\mathbf{r},t)\mathbf{H}^*(\mathbf{r},t)\right)\overset{\leftrightarrow}{\mathbf{I}}$$
$$-\varepsilon_0 \mathbf{E}(\mathbf{r},t)\mathbf{E}^*(\mathbf{r},t) - \mu_0 \mathbf{H}(\mathbf{r},t)\mathbf{H}^*(\mathbf{r},t), \quad (2)$$

where $\mathbf{E}(\mathbf{r},t)$ and $\mathbf{H}(\mathbf{r},t)$ are the electric and magnetic field, respectively, $\varepsilon_0$ and $\mu_0$ are the vacuum electric permittivity and magnetic permeability, respectively, and the symbol * denotes their complex conjugate, and $\overset{\leftrightarrow}{\mathbf{I}}$ is the identity tensor. The total electromagnetic force on the structure is given by integrating Eq. (1) over the whole volume $V$:

$$\mathbf{F}(\mathbf{r},t) = \int_V \left(\frac{d\mathbf{p}_{EM}(\mathbf{r},t)}{dt} + \frac{d\mathbf{p}_{ME}(\mathbf{r},t)}{dt}\right)dV = \int_V \nabla \cdot \overset{\leftrightarrow}{\mathbf{T}}(\mathbf{r},t)dV. \quad (3)$$

Therefore, the total electromagnetic force has two components: one related to the mechanical momentum and another to the electromagnetic momentum. However, for time-harmonic fields at a sinusoidal steady state, the time-averaged of electromagnetic momentum is zero, $\langle d\mathbf{p}_{EM}/dt\rangle = 0$, where $\langle \cdot \rangle$ denotes cycle averaging, thus $\langle \cdot \rangle = 1/2\, Re\{\cdot\}$. In other words, the mechanical frequency response of system is much slower than the electromagnetic fields oscillations at optical frequencies; thus,

$$\langle \mathbf{F}(\mathbf{r})\rangle = \int_V \left\langle\frac{d\mathbf{p}_{ME}(\mathbf{r})}{dt}\right\rangle dV = \int_V \left\langle\nabla \cdot \overset{\leftrightarrow}{\mathbf{T}}(\mathbf{r})\right\rangle dV. \quad (4)$$

Therefore, this method allows the rigorous calculation of the time-averaged total electromagnetic force acting on an object, by integrating the time-averaged MST, which is a function of the electromagnetic field spatial distributions, over a volume $V$ or, equivalently, over a surface $S$ enclosing the structure, by using the divergence theorem. This surface can be any arbitrary closed surface that includes the object volume embedded in vacuum [2]. Hence, since it depends on the integral of all six electromagnetic-field spatial distributions, for a given geometric parameter (gap) over an arbitrary inclosing surface, the MST method does not offer much physical insights about the optical force, besides that it may require a lot of computational effort [3,4,8]. Furthermore, the frequency dependence of the force is calculated by the Fourier transformation of the electromagnetic fields of the MST, which must be integrated over the frequency and over the surface.

On the other hand, from the first law of Thermodynamics, an adiabatic (lossless) system (null heat transfer, $Q = 0$), the changes in internal energy of the system, $U$ (potential, in this case), are opposite to the work done by the system, $W$, thus $dU = -W$ [12]. So, the time-average electromagnetic force on structure is given by

$$\langle \mathbf{F}(\mathbf{r}) \rangle = -\langle \nabla U_{\text{EM}}(\mathbf{r}) \rangle = \oint_S \langle \vec{\mathbf{T}}(\mathbf{r}) \rangle \cdot d\mathbf{S}. \quad (5)$$

Consequently, the time-averaged of the electromagnetic force on the lossless structure can be attributed either to the temporal variation of the time-averaged mechanical momentum or to the spatial variation of time-averaged electromagnetic energy.

Povinelli *et al.* have shown [1] that the (time-averaged) optical force $F_{\text{opt}}$ acting between two parts of a coupled dielectric system, separated by a generalized distance coordinate $q$, is given by:

$$F_{\text{opt}}(q) = -\frac{dU_{\text{opt}}(q)}{dq}\bigg|_{\mathbf{k}} = -\frac{d(N\hbar\omega(q))}{dq}\bigg|_{\mathbf{k}}$$
$$= -N\hbar \frac{d\omega(q)}{dq}\bigg|_{\mathbf{k}} = -\frac{U_{\text{opt}}(q)}{\omega(q)} \frac{d\omega(q)}{dq}\bigg|_{\mathbf{k}}, \quad (6)$$

where $U_{\text{opt}}$ is the optical (time-averaged) energy coupled into an eigenmode of the system, $N$ is the (average) eigenmode photon quantity, $\hbar$ is the Planck constant divide by $2\pi$, $\omega$ is the eigenmode angular frequency, and $\mathbf{k}$ is the eigenmode generalized wave vector; in this formalism, a negative optical force value corresponds to an attractive one. The ratio $g_{\text{OM}} = d\omega/dq$ is the optomechanical coupling coefficient, used as figure of merit of the system. This is a closed-system analysis, which assumes that an adiabatic change in a distance $dq$ will shift the eigeinmode frequency by $d\omega$. Notice that the derivative is taken at a fixed wave vector, to ensure the translational invariance of the optical mode and, therefore, that the photon number in this mode remains constant. This wave vector is general and can be cast as the Bloch wave vector in periodic systems (e.g. in photonic crystals). On the other hand, in simple coupled dielectric waveguides, the wave vector modulus of a guided mode is given by $|\mathbf{k}| = \beta = 2\pi/\lambda_0 \, n_{\text{eff}}$, where $\beta$ is the longitudinal propagation constant of the mode, $\lambda_0$ is the light source wavelength in vacuum, and $n_{\text{eff}}$ is the effective (refractive) index. The propagation constant ensues the invariance of the transverse field pattern of the guided mode in the longitudinal direction [13].

The optical power $P$ transmitted through the waveguide system of length $L$ is expressed as

$$P = \frac{d\omega(q)}{d\beta(q)} \frac{U_{\text{opt}}(q)}{L} = v_g(q) \frac{U_{\text{opt}}(q)}{L} = \frac{c}{n_g(q)} \frac{U_{\text{opt}}(q)}{L}, \quad (7)$$

where $v_g$ is the group velocity, $n_g$ is the group index, and $c$ is the speed of light in vacuum. Notice that Eq. (7) is obtained assuming linear lossless materials, but it takes into account the chromatic dispersion, which includes the material and the waveguide (guided) dispersions [13].

Substituting Eq. (7) in (6),

$$F_{\text{opt}}(q) = -\frac{U_{\text{opt}}(q)}{\omega(q)} \frac{d\omega(q)}{dq}\bigg|_\beta = -\frac{PL}{c} \frac{n_g(q)}{\omega(q)} \frac{d\omega(q)}{dq}\bigg|_\beta. \quad (8)$$

The optical force from the DR method represented in Eq. (8) is obtained by computing the eigenfrequency and group velocity of the guided mode at a fixed wave vector, for different separations values, and performing the derivative numerically. One important observation here it is that only some frequency-domain solvers are capable of keeping fix the wave vector, while calculating the eigenmode frequency, for a fixed source frequency. This equation agrees with the MST formalism, as shown in [1].

However, in most of the commercial photonic simulation softwares, based on eigenmode solvers, the eigenvalues of the modes are given in terms of the propagation constant $\beta$ or, equivalently, the effective index $n_{\text{eff}}$, rather than its respective eigenfrequency $\omega$; therefore, it is convenient to represent the optical force as a function of $n_{\text{eff}}$, by applying the following relation $\omega = c\beta/n_{\text{eff}}$. Eq. (8) then becomes [5]:

$$F_{\text{opt}}(q) = -\frac{PL}{c} \frac{n_g(q)}{n_{\text{eff}}(q)} \left( \frac{d\beta}{dq}\bigg|_\beta - \frac{dn_{\text{eff}}(q)}{dq}\bigg|_\beta \right)$$
$$= \frac{PL}{c} \frac{n_g(q)}{n_{\text{eff}}(q)} \frac{dn_{\text{eff}}(q)}{dq}\bigg|_\beta. \quad (9)$$

The optical force is attractive for $dn_{\text{eff}}/dq < 0$ and repulsive for $dn_{\text{eff}}/dq > 0$. The derivative in Eq. (9) is still taken at a fixed wave vector, as in Eq. (8).

Furthermore, by considering a monochromatic light with frequency $\omega_0$ or wavelength $\lambda_0$, since $\omega_0 = 2\pi c/\lambda_0$, it is possible to simplify Eq. (9) even more. By assuming an arbitrary function $f$ with three interdependent variables that obeys the following relation $f(\omega, \beta, q) = 0$, where each variable is an implicit function of the other two variables at $\omega_0$; the triple product (Euler's chain) rule states that

$$\frac{\partial \omega}{\partial \beta}\bigg|_q \frac{\partial \beta}{\partial q}\bigg|_{\omega_0} \frac{\partial q}{\partial \omega}\bigg|_\beta = -1, \quad (10)$$

and, by using the $v_g$ definition, shown in Eq. (7), we obtain:

$$\frac{\partial \omega}{\partial q}\bigg|_\beta = -v_g \frac{\partial \beta}{\partial q}\bigg|_{\omega_0} = -\frac{\omega}{n_g} \frac{\partial n_{\text{eff}}}{\partial q}\bigg|_{\omega_0}, \quad (11)$$

or, alternatively,

$$\frac{\partial n_{\text{eff}}}{\partial q}\bigg|_\beta = \frac{n_{\text{eff}}}{n_g} \frac{\partial n_{\text{eff}}}{\partial q}\bigg|_{\omega_0}. \quad (12)$$

Substituting Eq. (11) in Eq. (8), or Eq. (12) in Eq. (9), results in [6,7]:

$$F_{\text{opt}}(\omega_0, q) = \frac{PL}{c} \frac{dn_{\text{eff}}(\omega, q)}{dq}\bigg|_{\omega_0}. \quad (13)$$

Therefore, we obtain the optical force as a function of the effective index, with the derivative performed at a fixed light frequency, or wavelength. A graphical derivation of Eq. (13) from Eq. (9) is described in [7]. As a matter of fact, Eq. (12) is identical to equation 6 in [7], in the limit of infinitesimal variations of the gap. In summary, in Eq. (8) and (9), the derivatives must be performed at a fixed wave vector, and in Eq. (13) at a fixed frequency [7,8]. Recently, we applied the Minkowski Stress Tensor on coupled planar dielectric waveguides, where it is possible to obtain analytical expressions, in order to show that Eq. (13) may also be applied to different (non-vacuum) dielectric background media [14,15].

Rakich *et al*. used a different approach based on an open-system analysis to compute the optical forces - RTOF (*Response Theory of Optical Forces*) [9,10]. The RTOF method calculates the optical forces (and potentials) directly from the optical phase and amplitude responses of the optomechanical system. The RTOF method states that, for linear lossless device's materials, the optical force can be obtained by [9]

$$F_{\text{opt}}(\omega_0, q) = -\frac{\mathrm{d}U_{\text{opt}}(P, q)}{\mathrm{d}q} = \Phi(P)\hbar\frac{\mathrm{d}\phi(\omega_0, q)}{\mathrm{d}q}, \quad (14)$$

where $\Phi$ is photon flux through the waveguides, for a given optical power, and $\phi$ is the coordinate-dependent phase response of the system, which is dictated by the effective optical path length. For the coupled dielectric waveguides examined here, with a monochromatic light, they are given, respectively, by:

$$\Phi(P) = \frac{P}{\hbar\omega_0}, \quad \phi(\omega_0, q) = \frac{\omega_0}{c}n_{\text{eff}}(\omega_0, q)L. \quad (15)$$

By replacing the relations described in Eq. (15) into Eq. (14), we recover exactly the Eq. (13), as it has already been pointed out by [9], whose authors have also presented another derivation for Eq. (13), directly from Eq. (8), by using an analogy between the open and the closed system [9].

An important result obtained by RTOF method is that the optical forces are additive for distinct source frequencies and, therefore, multiple excitation frequencies in our system will lead a total optical force of the form:

$$F_{\text{opt}}^{\text{tot}}(\omega_i, q) = \sum_{i=1}^{n}\frac{P_i L}{c}\frac{\mathrm{d}n_{\text{eff}}(\omega_i, q)}{\mathrm{d}q}, \quad (16)$$

where $\omega_i$ and $P_i$ is the angular frequency and optical power from the $n - th$ light source. Eq. (16) can be transformed into a continuous integral to take into account any non-monochromatic, broadband or multiple optical source [9]. Furthermore, for a multimode waveguide under a monochromatic light excitation, the respective total optical force can be obtained using the RTOF multi-port analysis [10], yielding:

$$F_{\text{opt}}^{\text{tot}}(\omega, q) = \sum_{j=1}^{m}\frac{P_j L}{c}\frac{\mathrm{d}n_{\text{eff},j}(\omega, q)}{\mathrm{d}q}, \quad (17)$$

where $n_{\text{eff},j}$ and $P_j$ are the effective index and the optical power of the $j - th$ mode, respectively. The result of Eq. (17) can also be applied to the analysis of mode polarizations (quasi-TE and quasi-TM), when excited by unpolarized light [16]. Besides that, it can also be applied to the case of multiple degrees of freedom $q_n$, where $q_n = \{q_1, q_2, q_3, \ldots, n\}$ [10,16]. From Eqs. (16) and (17), one can see that, even in the more complex cases, these equations still resemble Eq. (13). The RTOF method also agrees with MST formalism [9,10].

Due to their computational simplicity, the two expressions based on derivative of the effective index have been largely applied to calculate the optical force in nano-optomechanical devices; for instance, Eq. (9) was used in [5, 17-32], and Eq. (13) in [6-9, 16, 33-46]. However, we notice that by applying Eq. (9) without the correct transformations, gives rise to an error on the magnitude of the optical force.

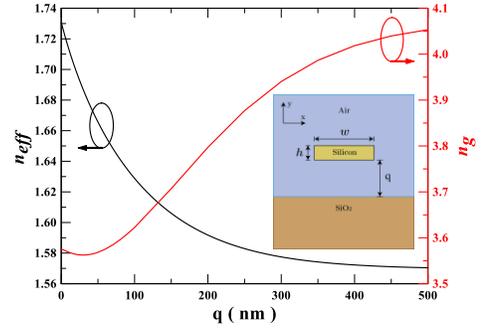

Fig. 1. Effective index and group index as a function of the gap distance between the Si nanowaveguide and SiO$_2$ substrate. The inset shows the transversal cross-section of the nano-optomechanical device.

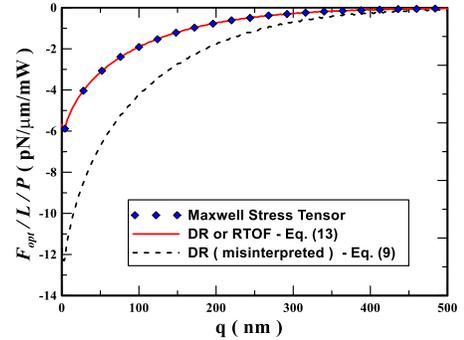

Fig. 2. Comparison between the normalized optical forces obtained through the MST method, the DR and RTOF method using the Eq. (13), and DR method using Eq. (9) on a commonly misinterpreted approach.

For example, we consider a typical nano-optomechanical device formed by a silicon (Si) waveguide (width $w = 500$ nm and height $h = 110$ nm) suspended over a silicon dioxide (SiO$_2$) substrate and surrounded by air, as illustrated on the inset in Fig. 1. We adopted the Si refractive index $n_{\text{Si}} = 3.5$, the SiO$_2$ refractive index $n_{\text{SiO2}} = 1.5$, and the air refractive index $n_{\text{Air}} = 1.0$, at the wavelength $\lambda_0 = 1.55$ μm, these particular values are chosen for consistency with [5, 26, 39, 47]. The dispersion diagram of the structure $n_{\text{eff}}$ and $n_g$, as a function of the gap distance $q$, shown in Fig. 1, was obtained for the fundamental (symmetric) quasi-TE eigenmode using a full-vectorial finite-difference mode solver. Then, the normalized optical force was calculated using the MST, DR or RTOF using Eq. (13), and the DR method was calculated using a commonly misinterpreted calculation of Eq. (9), as shown in Fig. 2.

Figure 2 shows that, when the gap is null, the optical force reaches a maximum absolute value of 6 pN/µm/mW, instead of 13 pN/µm/mW that is obtained by misinterpretation of Eq. (9). Therefore, Eq. (9), when applied without the correct transformation, overestimates the optical force by a factor of $n_g/n_{eff}$, since $n_g > n_{eff}$ in dielectric waveguides at non-anomalous frequency regions, as shown in Fig. 1; therefore, it disagrees with MST and with Eq. (13), obtained either by the DR or RTOF method. It is worth to stress that Eq. (9) is not incorrect, since it is just an intermediate step between Eq. (8) and Eq. (13), and should lead to the same correct results. The misinterpretation error is caused by employing the usual fixed wavelength approach, instead of ensuring a fixed wave vector, during the derivative of the effective index; such an error has been commonly observed in the technical literature [5, 26, 47]. In order to avoid it, one must apply a transformation, which turns out to be equivalent to multiply the Eq. (9) by a correction factor of $n_{eff}/n_g$, as show in Eq. (12), recovering exactly Eq. (13). We believe that this difference (factor $n_g/n_{eff}$) may be, in the near future, experimentally observed, especially for shorter separation [47].

In conclusion, we have applied first principles to show that the energy-based methods, such as DR and RTOF, must agree with the MST formalism in linear lossless systems. By using the RTOF method, we have also shown that the derivative of the effective index expression could be easily extended to more complex cases. Finally, we have used a literature example to show that the DR method, when expressed in terms of the derivative of the effective index, has to be used with caution, since a common misinterpretation can overestimate the absolute value of the optical force by a factor of $n_g/n_{eff}$.

**Funding.** Coordenação de Aperfeiçoamento de Pessoal de Nível Superior (CAPES); Conselho Nacional de Desenvolvimento Científico e Tecnológico (CNPq) (310855/2016-0 and 483116/2011)

**Acknowledgment**. J. R. Rodrigues was supported by CAPES and V. R. Almeida was supported by CAPES and CNPq.